\documentclass{iopart}
\usepackage{epsfig}
\newcommand{\beq}{\begin{equation}}
\newcommand{\eeq}{\end{equation}}
\newcommand{\bea}{\begin{eqnarray}}
\newcommand{\eea}{\end{eqnarray}}
\def\simlt{\buildrel < \over {_{\sim}}}
\def\simgt{\buildrel > \over {_{\sim}}}

\newcommand{\AmS}{{\protect\the\textfont2
  A\kern-.1667em\lower.5ex\hbox{M}\kern-.125emS}}

\begin{document}

\title{Phenomenological and Cosmological Implications of Neutrino Oscillations}

\author{S. F. King}

\address{Department of Physics and Astronomy, \\ 
        University of Southampton, Southampton SO17 1BJ, U.K.}

\begin{abstract}
I discuss the implications of neutrino oscillations for 
neutrino mass patterns and mass matrices, and for neutrinoless double beta
decay. I then go on to discuss the implications for 
cosmological relic density and dark matter, the 
galaxy structure limits on neutrino mass, nucleosynthesis, supernovae,
cosmic rays and gamma ray bursts.
\end{abstract}




\section{Introduction} 
Recent SNO results \cite{biller} when combined
with other solar neutrino data especially that of Super-Kamiokande
strongly favour the large mixing angle (LMA) MSW solar solution
\cite{choubey,Smirnov:2002in} 
with three active light neutrino states, and
$\theta_{12} \approx \pi/6$, 
$\Delta m_{21}^2\approx 5\times 10^{-5}{\rm eV}^2$.
The atmospheric neutrino data \cite{kajita} is consistent with
maximal $\nu_{\mu}- \nu_{\tau}$ neutrino mixing 
$\theta_{23} \approx \pi/4$
with $|\Delta m_{32}^2|\approx 2.5\times 10^{-3}{\rm eV}^2$
and the sign of $\Delta m_{32}^2$ undetermined. 
The CHOOZ experiment limits $\theta_{13} \simlt 0.2$
over the favoured atmospheric range \cite{Apollonio:1999ae}.

In this talk I shall briefly discuss some of the phenomenological and
cosmological implications of these results.

\section{Phenomenological Implications}
\subsection{Neutrino Masses and Mixing Angles}
The minimal neutrino sector required to account for the
atmospheric and solar neutrino oscillation data consists of
three light physical neutrinos with left-handed flavour eigenstates,
$\nu_e$, $\nu_\mu$, and $\nu_\tau$, defined to be those states
that share the same electroweak doublet as the left-handed
charged lepton mass eigenstates.
Within the framework of three--neutrino oscillations,
the neutrino flavor eigenstates $\nu_e$, $\nu_\mu$, and $\nu_\tau$ are
related to the neutrino mass eigenstates $\nu_1$, $\nu_2$, and $\nu_3$
with mass $m_1$, $m_2$, and $m_3$, respectively, by a $3\times3$ 
unitary matrix called the MNS matrix $U_{MNS}$
\cite{Maki:1962mu,Lee:1977qz}
\begin{equation}
\left(\begin{array}{c} \nu_e \\ \nu_\mu \\ \nu_\tau \end{array} \\ \right)=
\left(\begin{array}{ccc}
U_{e1} & U_{e2} & U_{e3} \\
U_{\mu1} & U_{\mu2} & U_{\mu3} \\
U_{\tau1} & U_{\tau2} & U_{\tau3} \\
\end{array}\right)
\left(\begin{array}{c} \nu_1 \\ \nu_2 \\ \nu_3 \end{array} \\ \right)
\; .
\end{equation}

Assuming the light neutrinos are Majorana,
$U_{MNS}$ can be parameterized in terms of three mixing angles
$\theta_{ij}$ and three complex phases $\delta_{ij}$.
A unitary matrix has six phases but three of them are removed 
by the phase symmetry of the charged lepton Dirac masses.
Since the neutrino masses are Majorana there is no additional
phase symmetry associated with them, unlike the case of quark
mixing where a further two phases may be removed.

\begin{figure}
\begin{center}
\epsfxsize=10cm
\epsfysize=5cm
\epsfbox{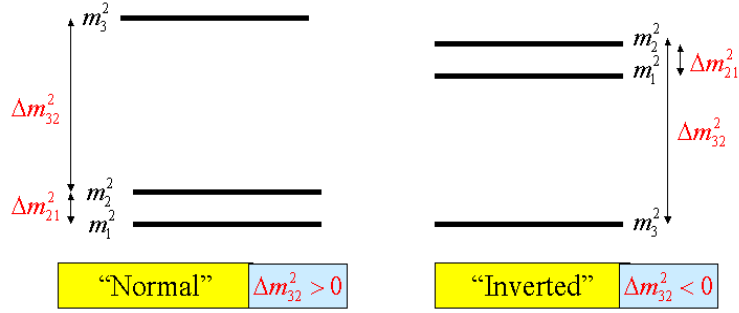}
\end{center}
\caption{\label{fig1} 
Alternative neutrino mass 
patterns that are consistent with neutrino
oscillation explanations of the atmospheric and solar data.}   
\end{figure}

There are basically two
patterns of neutrino mass squared orderings 
consistent with the atmospheric and solar data as shown in 
Fig.\ref{fig1}.


\begin{table*}[htb]
{\small
\caption{Leading order low energy neutrino Majorana mass matrices
$m_{LL}$ consistent with large atmospheric and solar mixing angles,
classified according to the rate of neutrinoless double beta decay
and the pattern of neutrino masses.}
\label{table1}
\newcommand{\m}{\hphantom{$-$}}
\newcommand{\cc}[1]{\multicolumn{1}{c}{#1}}
\renewcommand{\tabcolsep}{2pc} 
\renewcommand{\arraystretch}{1.2} 
\begin{tabular}{@{}|c|c|c|}
\hline
\hline
& Type I  & Type II \\ 
 & Small $\beta \beta_{0\nu}$ & Large $\beta \beta_{0\nu}$ \\
\hline
\hline
 & & \\ 
A & $\beta \beta_{0\nu}\simlt 0.0082$ eV & \\
Normal hierarchy  & & \\
$m_1^2,m_2^2\ll m_3^2$ & 
$\left(
\begin{array}{ccc}
0 & 0 & 0 \\
0 & 1 & 1\\
0 & 1 & 1 \\
\end{array}
\right)\frac{m}{2}$ & -- \\
& & \\
\hline
 & & \\ 
B & $\beta \beta_{0\nu}\simlt
0.0082$ eV & $\beta \beta_{0\nu}\simgt 0.0085$ eV\\
Inverted hierarchy & & \\
$m_1^2\approx m_2^2\gg m_3^2$ & 
$\left(
\begin{array}{ccc}
0 & 1 & 1 \\
1 & 0 & 0\\
1 & 0 & 0 \\
\end{array}
\right)\frac{m}{\sqrt{2}}$ & 
$\left(
\begin{array}{ccc}
1 & 0 & 0 \\
0 & \frac{1}{2} & \frac{1}{2}\\
0 & \frac{1}{2} & \frac{1}{2} \\
\end{array}
\right)m$
\\
& & \\
\hline                        
 & & \\ 
C &  &
$\beta \beta_{0\nu}\simgt 0.035$ eV \\
Approximate degeneracy & & diag(1,1,1)m\\
$m_1^2\approx m_2^2\approx m_3^2$ & 
$\left(
\begin{array}{ccc}
0 & \frac{1}{\sqrt{2}} & \frac{1}{\sqrt{2}} \\
\frac{1}{\sqrt{2}} & \frac{1}{2} & \frac{1}{2}\\
\frac{1}{\sqrt{2}} & \frac{1}{2} & \frac{1}{2} \\
\end{array}
\right)m$ & 
$\left(
\begin{array}{ccc}
1 & 0 & 0 \\
0 & 0 & 1\\
0 & 1 & 0 \\
\end{array}
\right)m$\\
& & \\
\hline
\hline                        
\end{tabular}\\[2pt]
}
\end{table*}


\subsection{Neutrino Mass Matrices}
In the flavour basis (in which the charged lepton mass matrix is diagonal),
for a given assumed form of $U_{MNS}$ and set of neutrino masses 
$m_i$ one may ``derive'' the form of the neutrino 
mass matrix $m_{LL}$, and this results in the candidate 
mass matrices in Table \ref{table1} \cite{Barbieri:1998mq,Altarelli:gu}.
In Table \ref{table1} the mass matrices are classified into two types:

Type I - small neutrinoless double beta decay

Type II - large neutrinoless double beta decay

They are also classified into the limiting cases consistent with the 
mass squared orderings in Fig.\ref{fig1}:

A - Normal hierarchy $m_1^2,m_2^2\ll m_3^2$

B - Inverted hierarchy $m_1^2 \approx m_2^2 \gg m_3^2$

C - Approximate degeneracy $m_1^2\approx m_2^2\approx m_3^2$

Thus according to our classification there is only one neutrino
mass matrix consistent with the normal neutrino mass hierarchy which 
we call Type IA, corresponding to the leading order neutrino masses
of the form $m_i=(0,0,m)$. For the inverted hierarchy there are two
cases, Type IB corresponding to $m_i=(m,-m,0)$ or Type IIB
corresponding to $m_i=(m,m,0)$. For the approximate degeneracy cases
there are three cases, Type IC correponding to $m_i=(m,-m,m)$
and two examples of Type IIC corresponding to either 
$m_i=(m,m,m)$ or $m_i=(m,m,-m)$. 

At present experiment allows any of the matrices in Table
\ref{table1}. In future it will be possible to uniquely specify the
neutrino matrix in the following way:

1. Neutrinoless double beta effectively measures the 11 element 
of the mass matrix $m_{LL}$ corresponding to  
\beq
\beta \beta_{0\nu}\equiv \sum_iU_{ei}^2m_i
\eeq
and is clearly capable of resolving Type I from Type II cases
according to the bounds given in Table \ref{table1} \cite{Pascoli:2002xq}.
There has been a recent claim of a signal in neutrinoless double 
beta decay correponding to $\beta \beta_{0\nu}=0.11-0.56$ eV at 95\% C.L.
\cite{Klapdor-Kleingrothaus:2001ke}.
However this claim has been criticised by two groups
\cite{Feruglio:2002af}, \cite{Aalseth:2002dt} and in turn this
criticism has been refuted \cite{Klapdor-Kleingrothaus:2002kf}. 
Since the Heidelberg-Moscow experiment has almost reached its full
sensitivity, we may have to wait for a next generation experiment
such as GENIUS which is capable 
of pushing down the sensitivity to 0.01 eV to resolve this question.

2. A neutrino factory will measure the sign of $\Delta m_{32}^2$
and resolve A from B.

3. Tritium beta decay experiments are sensitive to C
since they
measure the ``electron neutrino mass'' defined by
\beq
|m_{\nu_e}|\equiv \sum_i|U_{ei}|^2|m_i| \leq 2.2 {\rm eV} (95\% C.L.)
\label{tritium}
\eeq
where the current laboratory limit is shown \cite{Bonn:rz}.
The prospects for improving on this bound are good:
for example the KATRIN  experiment has a proposed sensitivity of 
0.35 eV. 

Model building based on the different neutrino mass matrices
in Table \ref{table1} was discussed separately at this meeting
\cite{graham} and will not be considered further here.

\section{Cosmological Implications}

\subsection{Cosmological relic density and dark matter}

In the early universe neutrinos were in thermal equilibrium with
photons, electrons and positrons. When the universe cooled to temperatures
of order 1 MeV the neutrinos decoupled, leading to a present day number
density similar to photons. If neutrinos have mass they contribute
to the mass density of the universe, leading to the constraint
\beq
\sum_i m_i\leq 28 \ {\rm eV}
\eeq
corresponding to 
\beq
\Omega_{\nu}h^2\leq 0.3
\eeq
in the standard notation where $\Omega_{\nu}$ is the ratio of 
the energy density in neutrinos to the critical energy density
of the universe (corresponding to a flat universe), 
and $h=H/100{\rm km}/{\rm Mpc}/{\rm s}$ is the scaled Hubble parameter.

From the current tritium limit on the elctron neutrino mass 
in Eq.\ref{tritium} together with the atmospheric and solar mass splittings
we have a stronger constraint
\beq
0.05 \ {\rm eV}\leq \sum_i m_i\leq 6.6 \ {\rm eV}
\eeq
corresponding to the range 
\beq
0.0005\leq \Omega_{\nu}h^2\leq 0.07
\eeq
which implies that neutrinos cannot be regarded as the dominant component
of dark matter, but make up between 0.1\% and 14\% of the mass-energy of the
universe.

\subsection{Galaxy structure limits on neutrino mass}

Recent results from the 2df galaxy redshift survey
indicate that 
\beq
\sum_i m_i <1.8 \ {\rm eV} (95\%C.L.)
\eeq
under certain mild assumptions \cite{Elgaroy:2002bi}.
This bound arises from the fact that the galaxy power spectrum
is reduced on small distance scales (large wavenumber) as the 
amount of hot dark matter increases. Thus
excessive amounts of free-streaming hot dark matter would
lead to galaxies being less clumped than they appear to be on
small distance scales.

However as discussed in \cite{Hannestad:2002iz} such bounds suffer
from parameter degeneracies due to a restricted parameter space,
and a more robust bound is 
\beq
\sum_i m_i <3 \ {\rm eV} (95\%C.L.)
\eeq
Combined with the solar and atmospheric oscillation data
this brackets the heaviest neutrino mass to be
in the approximate range 0.04-1 eV. The fact that the mass of the
heaviest neutrino is known to within an order of magnitude represents
remarkable progress in neutrino physics over recent years.
This implies that neutrinos make up less than about 7\%
of the total mass-energy of the universe, which is a stronger limit than
that quoted in the previous sub-section.

\subsection{Nucleosynthesis}

\begin{figure}
\begin{center}
\epsfxsize=7cm
\epsfysize=5cm
\epsfbox{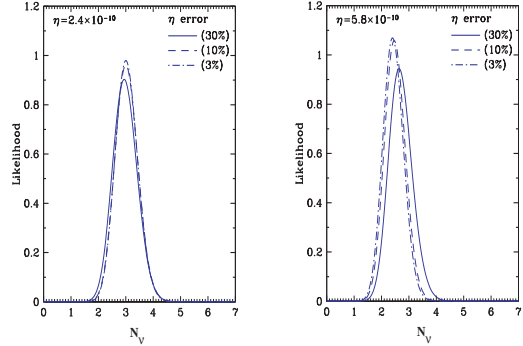}
\end{center}
\caption{\label{fig2} The distribution in the number of ``neutrino species''
(from \cite{Kainulainen:2002pu}).}   
\end{figure}

The number of light ``neutrino species'' $N_{\nu}$ (or any light species)
affects the freeze-out temperature of weak processes which determine the
neutron to proton ratio, and successful nucleosynthesis leads to 
the constraints on $N_{\nu}$ as shown in Figure \ref{fig2}.

\begin{figure}
\begin{center}
\epsfxsize=7cm
\epsfysize=5cm
\epsfbox{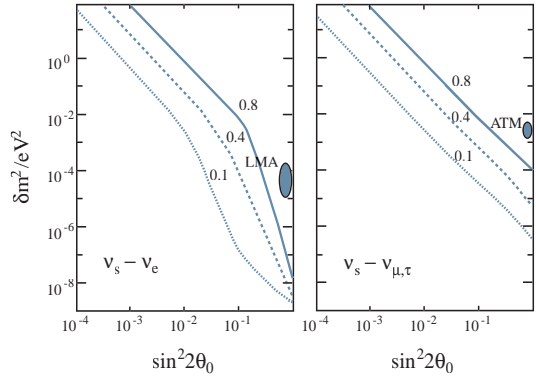}
\end{center}
\caption{\label{fig3} BBN constraints on active-sterile mixing parameters
(from \cite{Kainulainen:2002pu}).}   
\end{figure}

The limit on $N_{\nu}$ applies also to sterile neutrinos, but they
need to be produced during the time when nucleosynthesis was taking 
place, and the only way to produce them is via neutrino oscillations.
This leads to strong limits on sterile-active neutrino mixing angles
which tend to disfavour the LSND region, as shown in Figure \ref{fig3}.
Note that these limits assume that the lepton asymmetry is not
anomalously large \cite{Volkas:2000ei}.


\subsection{Supernovae}

It is well known that 99\% of the energy of a supernova is emitted
in the form of neutrinos, so one might hope that by studying supernovae
one can learn something about neutrino physics. Indeed studies of the
neutrino arrival times over a 10s interval from
SN1987A in the Small Magellanic Cloud have led to the 
mass bound \cite{Loredo:2001rx}
\beq
m_{\nu_e}\leq 6-20 {\rm eV}
\eeq
The LMA MSW parameter range may also have some observable
consequences for SN1987A \cite{valle,Minakata:2001cd}.

However it should be remarked that Supernova physics
has difficulties simulating explosions, and understanding how the spinning
neutron star remnant is kick-started.

\subsection{Cosmic Rays}

\begin{figure}
\begin{center}
\epsfxsize=5cm
\epsfysize=5cm
\epsfbox{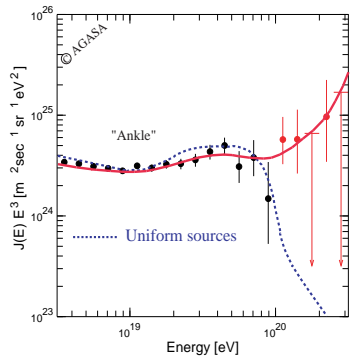}
\end{center}
\caption{\label{fig4} Scaled $E^3$ spectrum of cosmic rays. The solid
line is the prediction of the Z-burst model of \cite{Gelmini:2002xy}
(from \cite{Kainulainen:2002pu}).}   
\end{figure}

It has been a long standing puzzle why ultra high energy (UHE) cosmic rays
are apparently observed beyond the Greisen-Zatsepin-Kuzmin (GZK) cutoff
of $5\times 10^{19}$ GeV. The problem is that such cosmic rays
appear to arrive isotropically, which would imply an extragalactic
origin, whereas neither protons nor neutrons of such energies
could have originated from more than 50 Mpc away due to their scattering
off the cosmic background photons which would excite a delta resonance
if the energy exceeds $5\times 10^{19}$ GeV.
Yet despite this, several groups have reported handfuls of events
above the GZK cut-off, for example the AGASA data shown in 
Figure \ref{fig4}. Also shown in this figure is 
the prediction of a Z-burst model which was invented to explain
how the GZK cut-off could be overcome \cite{Gelmini:2002xy}. 
The basic idea of the Z-burst model is that UHE neutrinos
can travel much further, and scatter off the relic background
neutrinos to produce Z bosons within 50 Mpc.
Neutrino mass is required to increase the Z production rate, so
UHE cosmic rays and neutrino mass are linked in this model.

Reecently data from Fly's Eye, Yakutsk and expecially HiRes
have brought into question whether the GZK cut-off is being exceeded
at all. A recent analysis based on these data
suggests that there might be no problem with the GZK cut-off
\cite{Bahcall:2002wi}. On the other hand Fly's Eye and HiRes establish
the cosmic ray energy from observing the fluorescence of Nitrogen
molecules using a rather dated technique.
The issue should be resolved soon by the Pierre Auger
Observatory, which can both measure the shower's pattern on
the Earth's surface, and the fluorescence of the shower in the air.


\begin{figure}
\begin{center}
\epsfxsize=5cm
\epsfysize=5cm
\epsfbox{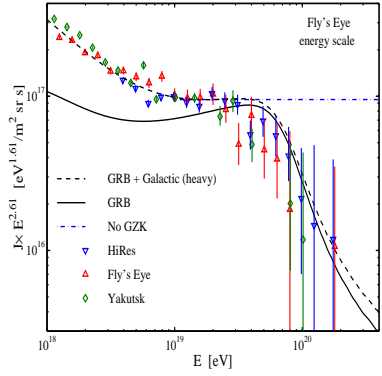}
\end{center}
\caption{\label{fig5} Recent data on cosmic rays
(from \cite{Bahcall:2002wi}) which shows that the GZK cut-off
might be respected after all.}   
\end{figure}


\begin{figure}
\begin{center}
\epsfxsize=10cm
\epsfysize=7cm
\epsfbox{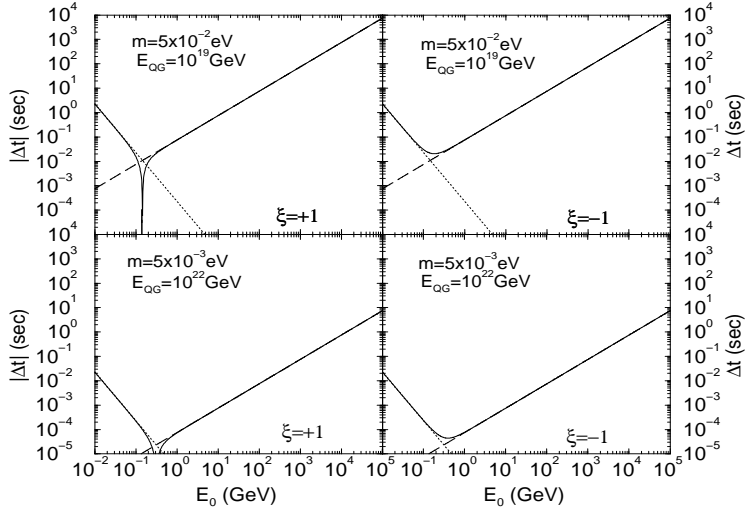}
\end{center}
\caption{\label{fig6} Time delays plotted against the neutrino energy.
The downward sloping solid curve in each panel results from 
the effect of neutrino mass dominating at low energy, while the
upward sloping solid curves are due to quantum gravity effects
dominating at high energy (from \cite{Choubey:2002bh}).}   
\end{figure}


\subsection{Gamma Ray Bursts (GRBs)}

GRBs are distant ($z\sim 1$), energetic and enigmatic.
According to the fireball model (for a review see \cite{Ghisellini:2001jr})
GRBs may result from the core collapse of a very massive supernova
to a compact, rotating black hole. The energy is emitted in beamed
relativistic fireball jets which are expected to contain copious
neutrino fluxes. GRBs can thus be regarded as a high intensity, high
energy neutrino beam with a cosmological baseline.

In some models of quantum gravity \cite{Ellis:1999sf} neutrinos of
mass $m$ obey the dispersion relation
\beq
p^2c^2\approx E^2(1-\xi\frac{E}{E_{QG}})-m^2c^4
\eeq
where $E_{QG}$ is the scale of quantum gravity. A consequence
of this dispersion relation is that the effects of both quantum
gravity and neutrino mass will cause time delays in the arrival
times of neutrinos compared to low energy massless photons,
as shown in Figure \ref{fig6}. Using the millisecond time
structure of GRBs it may be possible to measure neutrino masses 
from time of flight delays, in principle down to
0.001 eV \cite{Choubey:2002bh}. However this estimate assumes
complete understanding of GRBs, and large enough event rates,
and in practice there are many theoretical and experimental
challenges to be overcome.

\section{Conclusions}

Future neutrino oscillation experiments, will give precious 
information about the mass squared splittings  
$\Delta m_{ij}^2\equiv m_i^2-m_j^2$, mixing angles, and 
CP violating phase. KATRIN and GENIUS will tell us about the
absolute scale and nature of neutrino mass. We have seen that
cosmology and astrophysics
is also sensitive to the absolute values of neutrino masses.
In the coming years there should be a fascinating interplay between
neutrino physics on Earth and in the Heavens.


\end{document}